\documentstyle[graphicx,multicol,pre,aps]{revtex}

\begin{document}

\draft

\title{Nonlinearity-induced conformational instability and dynamics 
of biopolymers}

\author{Serge F. Mingaleev$^{1,2,3}$,
Yuri B. Gaididei$^{2,3}$, Peter L. Christiansen$^{2}$,
and Yuri S. Kivshar$^{1}$}
\address{$^{1}$ Nonlinear Physics Group,  
Research School of Physical Sciences and Engineering, \\
Australian National University, Canberra ACT 0200, Australia \\
$^{2}$ Department of Mathematical Modeling,
The Technical University of Denmark, DK-2800 Lyngby, Denmark \\
$^{3}$ Bogolyubov Institute for Theoretical Physics, 14-B Metrologichna
St., 03143 Kiev, Ukraine}

\maketitle

\begin{abstract}
We propose a simple phenomenological model for describing the
conformational dynamics of biopolymers via the nonlinearity-induced
buckling and collapse (i.e. coiling up) instabilities. 
We describe the buckling instability analytically, and then
demonstrate the role of both instabilities 
in the folding of biopolymers through the 
numerical simulations of the three-dimensional dynamics of 
a long semiflexible chain in the aqueous environment. 
\end{abstract}


\begin{multicols}{2}
\narrowtext


Conformational flexibility is a fundamental property of polymers 
which differentiates them from small molecules and gives rise to
their remarkable properties. 
However, even the properties of polymers seem meagre when compared 
to the functionality of biopolymers. 
A distinctive feature of biopolymers is that, on the one hand, 
they are heterogeneous and, on the other, their elementary sub-units 
have a complex structure and can carry long-lived nonlinear 
excitations \cite{myoglobin,peyrard,davydov,dna,scott,cruzeiro}. 
While the significance of the heterogeneity gained 
wide recognition as being of fundamental importance 
for protein folding \cite{Taylor:2001-apr:RPP}, 
the function of nonlinear excitations remains obscured, 
in spite of a large body of theoretical research 
\cite{peyrard,davydov,dna,scott}. 

Here we suggest {\em a new role of nonlinear 
excitations} in one of the most important functionalities of 
biopolymers, their {\em conformational dynamics}. 
For the first time to our knowledge, we demonstrate 
that nonlinear excitations 
can facilitate bending or folding of a semiflexible 
molecular chain via the buckling or collapse 
instability, and thus provide a 
possible scenario for the conformational dynamics of 
biopolymers.
We consider a semiflexible chain that models a biopolymer and 
assume that its mechanical degrees of freedom are coupled to the
internal degrees of freedom such as amide-I vibrations 
in proteins \cite{myoglobin,peyrard,davydov}, base-pair 
vibrations in DNA \cite{peyrard,dna}, or polarons in 
proteins \cite{peyrard,davydov}. 
This coupling provides the energy needed to
overcome the potential barriers for folding, thus avoiding the need
for stochastic thermal activation. 
We reveal that such a simple model describes the conformational
dynamics as {\em a deterministic controlled process}: when the
amplitude of the internal excitation becomes large, the
chain displays the buckling or collapse instability that
subsequently initiates the chain folding.
We believe that such simple physics can also account for the
kinetics of conformational phase transitions of semiflexible
polymers in solutions \cite{geom}.

The importance of nonlinear excitations has already been 
emphasized in the modeling of biopolymers and conjugated 
polymers. In particular,
it was suggested that solitons may provide a possible physical mechanism
for the energy (or charge) transport and storage in proteins 
\cite{peyrard,davydov}
and conducting  polymers \cite{polymers}. The concept of nonlinear
localized modes was also employed to explain some
specific features of the DNA dynamics \cite{peyrard,dna}. 
However, most of those
studies were dealing with nonlinear properties of  {\em straight} 
molecular chains and never discussed changes in their geometrical
structure. The study of nonlinear properties of {\em curved chains} has been
initiated only recently \cite{curv} and, in particular,
it was shown that the bending in a curved chain manifests 
itself as an effective trap for nonlinear localized modes. 
In this Letter, we make an important step forward and, 
taking into account {\em flexibility} of biopolymer chains, 
demonstrate that nonlinear excitations may act as drivers 
giving impetus to {\em conformational dynamics} of 
biopolymers.


Let us consider a simple phenomenological model of a biopolymer 
consisting of particles with mass $M$, indexed by $n$, and located at the
points 
$\bbox{r}_n=\{x_n, \, y_n, \, z_n\}$.  We assume that each particle
represents a complex sub-unit of the biopolymer which, 
additionally to its position $\bbox{r}_n(t)$, 
carries an internal excitation which can 
be characterized, in some approximation 
\cite{peyrard,davydov,dna,scott,polymers}, 
by the complex amplitude, 
$\psi_n(t)$. Such an internal mode
can represent, for instance, a polaron state 
(an excess electron accompanied by phonons) or vibrational state
(amide-I vibrations in proteins or base-pair vibrations in DNA).
The Hamiltonian of such a chain is written as
$H= T + U + V$, where 
$T= (M/2) \sum_{n} (d \bbox{r}_n/dt)^2$ is the kinetic energy, 
$U(\bbox{r}_n)$ is the potential energy
of inter-particle interactions, and 
\begin{displaymath}
V(\psi_n, \bbox{r}_n)= \sum_{n} \left\{ 2|\psi_n|^2 -
\sum_{m \neq n} J_{n m} \psi^*_n \psi_{m} -
\frac{1}{2} \,\chi\, |\psi_n|^4 \right\} 
\end{displaymath}
is the energy of the internal excitation.
Here $\chi$ characterizes on-site self-trapping 
nonlinearity of the internal excitation, and 
$J_{nm}$ are the excitation transfer coefficients that 
depend on the distance in the embedding space  between the 
particles $n$ and $m$: 
$J_{n m} \equiv J(|\bbox{r}_n -\bbox{r}_m|)= (e^{\alpha}-1) 
\exp\left(-\alpha |\bbox{r}_n-\bbox{r}_{m}|\right)$ 
\cite{note0}. 
From the Hamiltonian, we obtain the equations of motion 
\begin{eqnarray}
M \frac{d^2 \bbox{r}_n}{dt^2} + \nu \frac{d \bbox{r}_n}{dt} + 
\frac{d U}{d \bbox{r}_n} - \sum_{k} \sum_{m \neq k}
\frac{d J_{k m}}{d \bbox{r}_n} \psi^*_k \psi_{m} = 0 \; , 
\nonumber \\
i \frac{\partial \psi_n}{\partial t} - 2 \psi_n +
\sum_{m \neq n} J_{nm} \psi_m + \chi \, |\psi_n|^2 \psi_n =0 \; ,
\label{eq-mot}
\end{eqnarray}
where we allowed additionally for a viscous damping $\nu$ of the 
aqueous environment. We assume that the number of excitations, 
$\sum_{n} |\psi_n|^2=1$, is conserved by Eq. (\ref{eq-mot}). 
This approximation is fully justified for polaron states 
and should be fairly good for vibrational states, 
taking into account that in the case under study the lifetime 
of amide I vibrations is greatly inceased as a result of their 
self-trapping localization \cite{myoglobin,cruzeiro}.

\begin{figure}
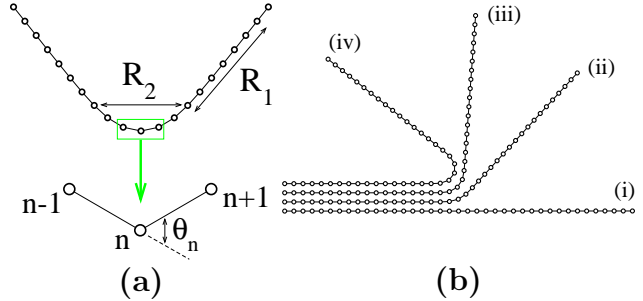

\begin{center}
\begin{minipage}{35mm}
\includegraphics[width=35mm,angle=0,clip]{bio-fig1a.eps}
\centerline{{\large\bf (a)}}
\end{minipage}
\begin{minipage}{47mm}
\includegraphics[width=47mm,angle=0,clip]{bio-fig1b.eps} \\[6mm]
\centerline{{\large\bf (b)}}
\end{minipage}
\end{center}
\caption{(a) Schematic representation 
of a buckled chain.  The distance between the sites
near the bend becomes smaller  ($R_2<R_1$) lowering 
the coupling energy of a nonlinear 
mode that works against the increased bending energy and lead to 
a buckling instability. (b) Ground state spatial configurations 
of a semiflexible chain  ($\alpha=2$, $\theta_{\rm max}=\pi/3$,
and $\kappa=0.125$) at different values of $\chi$: 
2.13~(i), 2.26~(ii), 2.44~(iii), 2.88~(iv), 3.00~(iv), 
3.13~(iii), 3.18~(ii), and 3.25~(i). As the nonlinearity $\chi$ 
increases, the chain bending initially increases, reaches its 
maximum at $\chi\approx 2.96$, and then decreases.}
\label{fig:chain}
\end{figure}

We consider the potential energy of inter-particle interactions as a sum, 
$U(\bbox{r}_n)=U_{S}+U_{B}+U_{R}$, of the stretching energy 
$U_{S} = (\sigma/2) \sum_{n} (|\bbox{r}_n-\bbox{r}_{n-1}|-a)^2$, 
bending energy \cite{note2} 
\begin{equation}
U_{B} = \frac{\kappa}{2} \sum_{n}
\frac{\theta_n^2}{1-(\theta_n / \theta_{\rm max})^2} \; , 
\label{poten-UB}
\end{equation}
and (to describe correctly coiling up of the chain) the 
energy of short-range repulsive interactions between particles 
(considered as elastic balls of the diameter $d$): 
\begin{equation}
U_{R} = \frac{\delta}{2} \sum_{n} \sum_{m \neq n}
(d-|\bbox{r}_n-\bbox{r}_{m}|)^2 \; ,
\label{poten-UR}
\end{equation}
for $|\bbox{r}_n-\bbox{r}_{m}|<d$, and $U_{R} = 0$,
otherwise.
Here $a$ is the equilibrium lattice spacing (we take $a=1$), 
$\sigma$ is the elastic modulus of the stretching 
rigidity of the chain, $\theta_n$ is the angle 
between neighboring bond vectors, 
$(\bbox{r}_{n} - \bbox{r}_{n-1})$ and $(\bbox{r}_{n+1} - \bbox{r}_{n})$, 
that meet at the $n$-th particle [see Fig. \ref{fig:chain}(a)],
$\theta_{\rm max}$ is a maximum bending angle, and 
$\kappa$ is the elastic modulus of the bending rigidity of the chain. 

First of all, we recall the properties 
of nonlinear localized modes in a curved chain, when the 
chain geometry is ``frozen''. Such an analysis has been 
recently carried out in Ref. \onlinecite{curv}, where it was shown 
that the chain curvature affects strongly the properties of 
nonlinear modes, creating an effective double-well potential around 
the bending point, which 
is responsible for the symmetry-breaking effect: 
as the nonlinearity $\chi$ increases above some threshold, 
a symmetric 
stationary mode becomes unstable, and it transforms into
an energetically more favorable asymmetric stationary mode.
In that study, it was also found that the energy of a 
nonlinear mode decreases as a square-law of the curvature 
when the curvature is small. 
Thus, since the bending energy of the chain increases
also as the square of the curvature, one can expect that
there should exist a converse influence of the internal nonlinear
modes on the conformational dynamics of the chain. 
Below, we confirm this conjecture by straightforward 
numerical calculations and find that not only the chain tends
to bend, for small enough $\kappa$, but it also 
collapses (i.e. coils up), for smaller $\kappa$.


As the first step of our study in this Letter, we analyze 
the stationary ground-state configurations of 
the chain described by the model introduced above. 
We assume that the chain is infinitely long, 
planar ($z_n = 0$), and inextensible ($\sigma \to \infty$),
so that the distance between  the neighboring sites of the chain remains
constant. In this case,  a  spatial conformation of the chain is 
completely defined by the bending angles $\theta_n$.

\begin{figure}
\centerline{\hbox{
\includegraphics[width=70mm,angle=0,clip]{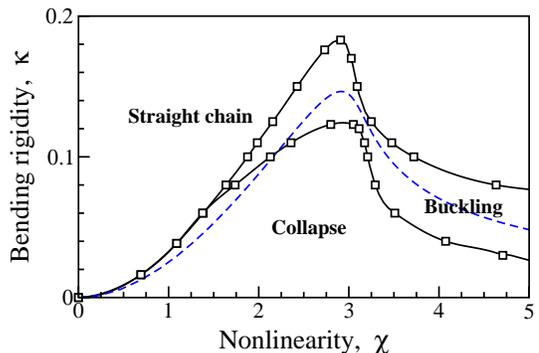}}}
\caption{Phase diagram of the 
ground states of a semiflexible biopolymer chain calculated 
numerically (solid curves with squares) and analytically 
from Eqs.~(\protect\ref{kappa-cr})--(\ref{kin}) (dashed curve), at 
$\alpha=2$, $\theta_{\rm max}=\pi/3$, and $\sigma=\infty$.}
\label{fig:kappa-lambda}
\end{figure}


We find the ground-state configurations of
the chain looking for the stationary solutions
of Eq. (\ref{eq-mot}) in the form
$\psi_n(t) = e^{i \Lambda t} \phi_n$ and 
$d \bbox{r}_n/dt = 0$. The corresponding 
system of nonlinear algebraic equations
has been solved by the Newton-Raphson iteration scheme.
The results of these calculations are presented 
in Fig. \ref{fig:kappa-lambda}. The most interesting feature of this
phase diagram is the existence of {\em three distinct 
types} of the ground states.

\begin{figure}
\centerline{\hbox{
\includegraphics[width=73mm,angle=0,clip]{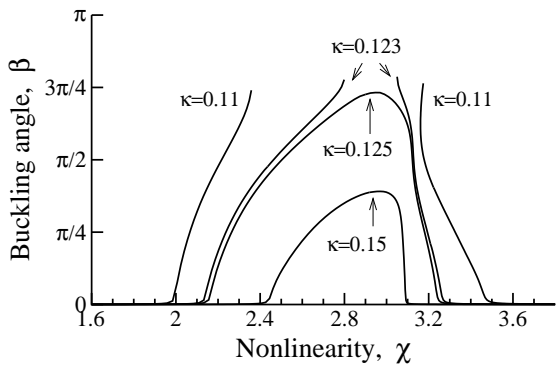}}}
\caption{Buckling angle
$\beta$ vs. the nonlinearity $\chi$ for 
$\alpha=2$, $\theta_{max}=\pi/3$, $\sigma=\infty$, 
and varying $\kappa$.}
\label{fig:beta-lambda}
\end{figure}

A stiff chain (with $\kappa>0.183$) is always straight 
and, regardless of the value of $\chi$,  it cannot 
be affected by the presence of a nonlinear excitation. 
However, for $\kappa<0.183$, there appears a finite domain 
of $\chi$, for which a straight chain becomes unstable 
in the presence of a nonlinear mode. Inside of this domain, 
the chain in its ground state {\em curves symmetrically} 
around the nonlinear excitation, with some finite overall buckling 
angle $\beta=\sum_n \theta_{n}$ [see Fig. \ref{fig:chain}(b)].
In Fig. \ref{fig:beta-lambda}, we plot the dependence of
the buckling angle $\beta$ vs. $\chi$ for 
different values of~$\kappa$. As is seen, for $0.124<\kappa<0.183$ the 
buckling angle $\beta$ is always finite and 
reaches its maximum at $\chi \simeq 3$.

\begin{figure}
\begin{center}
\includegraphics[width=78mm,angle=0,clip]{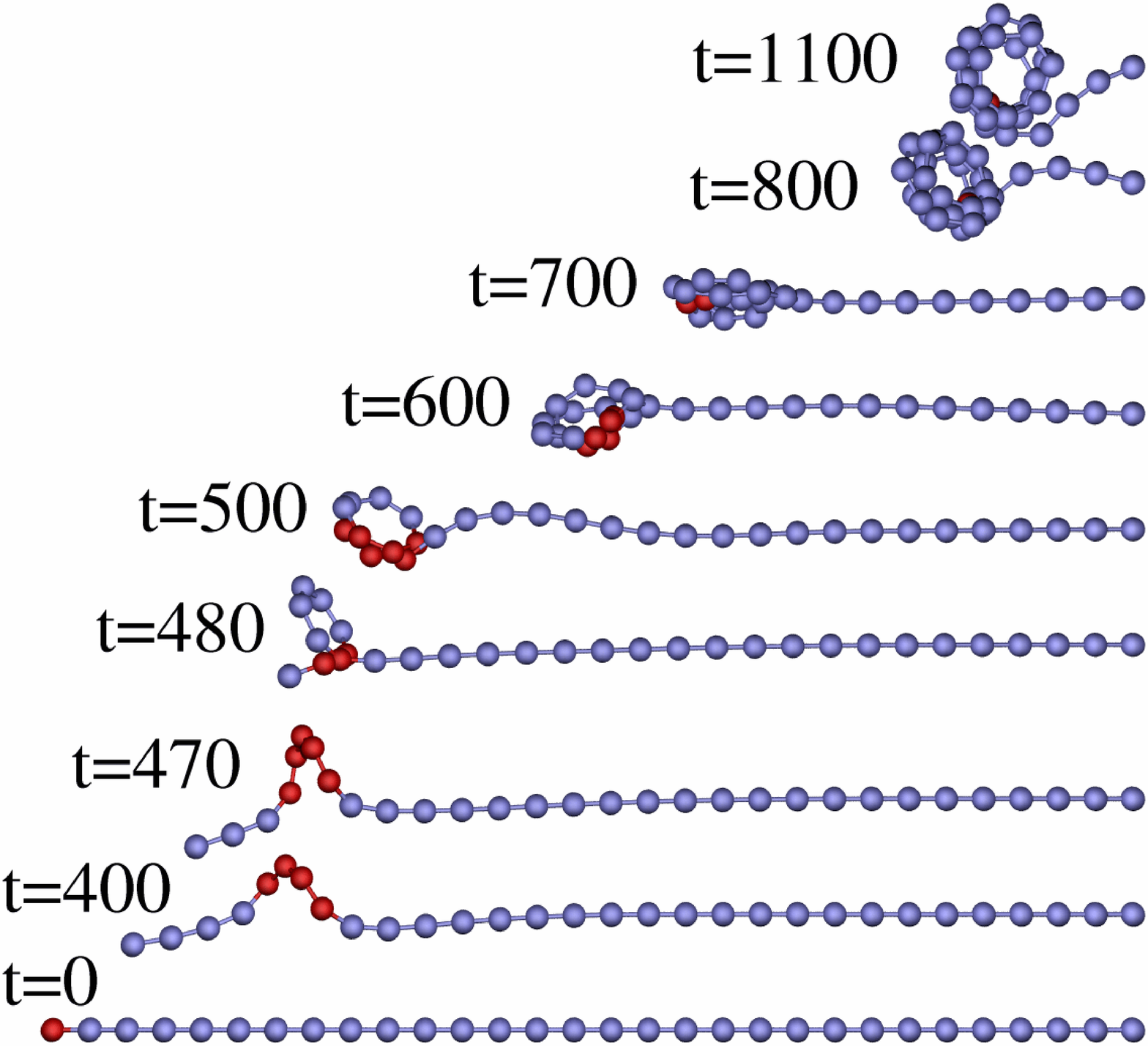}
\end{center}
\caption{Dynamics of a semiflexible biopolymer chain due to 
the collapse instability 
produced by a moving nonlinear mode generated from the 
single-particle excitation at the left end of the chain. 
The chain consists of 30 particles, with the right end fixed. 
Marked particles indicate location of the moving excitation 
(for these particles, $|\psi_n|^2>0.25$).
The parameters are: $\chi=3.2$, $\kappa=0.06$, $\nu=0.3$, 
$\sigma=10^3$,  $\alpha=2$, 
$\theta_{\rm max}=\pi/3$, $M=0.5$, $\delta=10^2$, and $d=0.6$.}
\label{fig:dyn}
\end{figure}

When the bending rigidity of the chain decreases further,
i.e. for $\kappa<0.124$, a new type of the ground state,
{\em a collapsed chain}, emerges 
(see Figs. \ref{fig:kappa-lambda} and
\ref{fig:beta-lambda}). In this case, the distance
between the complementary particles of the chain 
[see Fig. \ref{fig:chain}(b)] decreases 
to an extent that the growing interaction  between the tails
of the nonlinear excitation turns buckling into collapse 
causing the coiling up of the chain. 
The conformation of the chain in the collapsed state 
(which is usually a three-dimensional compact globule, 
see Fig. \ref{fig:dyn}) 
is sensitive to the peculiarities of the potentials 
(\ref{poten-UB})--(\ref{poten-UR}) and initial conditions.

To gain a better insight into the physical 
mechanism of the buckling instability,
we have developed a simple variational approach. 
We assume that the chain bending occurs only at the center of the 
excitation ($n=0$). In this case, the chain is composed of two
semi-infinite lines with the overall buckling angle  $\beta \equiv
\theta_{0}$
between them. Obviously, such a geometry cannot describe the chain collapse 
because $\beta$ is restricted by  the maximum angle $\theta_{\rm
max}$,  but it should be valid as long as we are not interested in the
collapsed
phase but study only a threshold between the straight and buckled 
chain phases.

In such a simplified model, the ground state of the chain is determined
by the dependence of the system energy on $\beta$. 
We use a variational ansatz for the localized mode in the form
$\phi_n=\sqrt{\tanh \gamma\,} \, \exp(-\gamma \, |n|)$, 
which satisfies the constraint  $\sum_n \phi_n^2=1$, and we minimize $H$ 
with respect to $\beta$ and $\gamma$. 
Thus, we can show that the straight configuration of the chain 
($\beta=0$) corresponds to the energy maximum (i.e., the straight 
chain is unstable) for $\kappa\,<\kappa_{\rm cr}$,
where the critical value of the bending stiffness
is given by the following analytical expression 
\begin{eqnarray}
\label{kappa-cr}
\kappa_{\rm cr}=\chi \, \alpha (e^{\alpha}-1)
\frac{3\,e^{\alpha+\gamma}-1}{3 \,
(e^{\alpha+\gamma}-1)^3} \, \tanh \gamma \; ,
\end{eqnarray}
and $\gamma$ is determined from the equations
\begin{eqnarray}
\label{kin}
\chi = \frac{8 \cosh^4\gamma}{(4 \cosh^2\gamma-3)}\,
\frac{\partial Q(\gamma)}{\partial \gamma} \; , 
\nonumber \\
Q(\gamma) = \frac{e^{\alpha} \, (e^{\gamma}-1)^2
(e^{\alpha+2\gamma} +1)}{(e^{2\gamma}+1) \,
(e^{\alpha+\gamma}-1)^2} \; .
\end{eqnarray}
In Fig. \ref{fig:kappa-lambda}, we compare the
threshold curve that separates the phases of the 
straight and buckled conformations 
predicted by the analytical formula 
(\ref{kappa-cr})--(\ref{kin}), 
with the same dependence calculated numerically. 
One can see that the variational method provides
reasonably fair analytical results for the phase diagram.


An important question is how the buckling and
collapse instabilities manifest themselves in the 
dynamics of semiflexible biopolymer chains. To clarify this issue,
we consider a simple (but physically important) case of an 
initially straight chain of a finite  length whose left
end is free whereas the right end is fixed. 
First, we excite a single particle at the left end of the chain, and
then keep track of how this excitation propagates along the chain
and how it modifies the chain geometry. For small values 
of $\chi$, such a single-particle excitation disperses and  spreads out,
whereas for very large values of $\chi$ it does not propagate
and remains completely trapped at the end of the chain. 
However, for the intermediate values of $\chi$ (in our model, for 
$2.5 < \chi <3.5$), such a single-particle drive applied at the end of 
the chain generates {\em a moving localized mode}. 
It is remarkable that this interval of nonlinearity 
parameter $\chi$ matches closely the region in which 
the bending capability of the nonlinear mode 
reaches its maximum.


For a stiff biopolymer chain (i.e., for  $\kappa > 0.183$), 
the propagation of the generated nonlinear mode does not produce
any conformational change of the chain.  However, when $\kappa$ 
decreases, the nonlinear mode generates a local chain bending. 
In this case the nonlinear mode propagates
along the chain {\em being accompanied by a local  bending};  
the amplitude of such a bending decreases as the viscous damping $\nu$ 
increases. Thus, the buckling instability
we have described above 
appears to be physically visible, and it is responsible, in particular,
for large-amplitude  localized bending waves. 


Let us consider now a flexible biopolymer chain with even 
smaller $\kappa$, for which the nonlinear mode should produce a collapse
instability (see Fig. \ref{fig:kappa-lambda}). In this case, the 
evolution of the chain
clearly demonstrates a collapse dynamics (see Fig.
\ref{fig:dyn}), and the chain folds 
into a compact coil with 
several loops consisting of 7--10 particles each
\cite{note1}.
Importantly, even for large values of the damping parameter $\nu$ 
the collapse of the chain is only delayed, but not prevented. 
For instance, for the parameters used in 
Fig. \ref{fig:dyn} the collapse-on time $t_c$ grows almost 
linearly with $\nu$: $t_c \approx 87 + 1290 \nu + 283 \nu^2$ 
for $\nu \leq 1$. Besides, increasing of $\nu$ leads to 
decreasing of the distance between the place of the collapse 
nucleation and the left end of the chain. 


It should be mentioned that the instability 
we discuss in this Letter remains latent 
in a straight infinitely long chain, because the
bending of such a chain 
would require an infinite energy.  However, as we have 
demonstrated above, the instability
manifests itself as soon as we consider more realistic cases and
take into account a finite length of the chain.  
In this case, a nonlinear
localized mode, once generated at the end of the chain, 
propagates along the chain 
{\em accompanied by a local chain bending}, for 
intermediate values of the bending rigidity $\kappa$, 
and it {\em causes the chain folding}, 
for smaller $\kappa$. 

In conclusion, we have predicted analytically 
and demonstrated numerically a new role played by nonlinear 
excitations in the conformational 
dynamics of biopolymers. Taking into account
the coupling between the internal and mechanical degrees of freedom of
a semiflexible biopolymer chain, we have found that it may produce 
the buckling and collapse instabilities of an initially straight chain. 
In the case of collapse instability, the viscous damping of the aqueous 
environment only slows down the folding (coiling up) of the chain, 
but does not stop it even for relatively large damping ($\nu \simeq 1$). 
It should be stressed that these 
instabilities are most pronounced for {\em intermediate values} of the 
self-trapping nonlinearity, and they vanish completely 
in the linear limit (i.e., for $\chi=0$). 
We find that this effect is only weakly affected by 
the peculiarities of the interaction potentials \cite{note0,note2}, 
and thus it should be generic for different models of nonlinear 
semiflexible chains. 

We believe that the nonlinearity-induced buckling 
and collapse instabilities may be of a considerable 
importance in the conformational dynamics of the 
globular macromolecules 
with a multitude of quasi-isoenergetic conformations 
\cite{Taylor:2001-apr:RPP}, 
where nonlinear modes excited even far from the chain ends 
should cause {\em conformational transitions} between 
different types of the globular states of the same or similar topology.
Presumably, another related problem is the energy 
transduction in molecular motors, for which it was recently 
suggested that the free energy 
from ATP hydrolysis is initially converted into the 
hinge-bending motion of the motor sub-unit 
\cite{Wang:1998:Nature}. 


The authors are indebted to M.~Peyrard, 
K.~Rasmussen, S.~Takeno, 
J.~Tuszynski, and A.~Zolotaryuk for useful and 
encouraging comments. 
S.~M. and Yu.~G. acknowledge support from the European 
Commission RTN project LOCNET (HPRN-CT-1999-00163) and the 
MIDIT center at the Technical University of Denmark, S.~M. 
acknowledges supports from the the Performance and Planning 
Fund of the Institute for Advanced Studies, the Australian 
National University, and Yu.~G. from the DLR Project No. 
UKR-002-99 and the SRC ``Vidhuk'' (Kiev).



\end{multicols}

\begin{thebibliography}{99}
\bibitem{note0}
Discussed effects are tolerant to 
a choice of $J_{nm}$, provided the interaction extends 
beyond the approximation of the nearest neighbors. 
We have verified the validity of our predictions for 
other types of $J_{nm}$, in particular, for the 
dipole-dipole interaction.
\bibitem{note2}
We have checked that the use of other types of the 
bending potential leads to qualitatively same results.
\bibitem{note1}
We use $\theta_{\rm max}=\pi/3$, so
that 7 is a minimum number of particles in such 
a loop.
\end{thebibliography}
\end{document}